\documentclass[runningheads]{llncs}
\usepackage[T1]{fontenc}
\usepackage{graphicx}
\usepackage{framed,enumitem} 
\usepackage[dvipsnames]{xcolor}
\begin{document}
\title{NFDIcore 2.0: A BFO-Compliant Ontology for Multi-Domain Research Infrastructures}

\author{Oleksandra Bruns\inst{1,2}\orcidID{0000-0002-8501-6700} \and
Tabea Tietz\inst{1,2}\orcidID{0000-0002-1648-1684} \and 
Jörg Waitelonis\inst{1,2}\orcidID{0000-0001-7192-7143} \and 
Etienne Posthumus\inst{1}\orcidID{0000-0002-0006-7542} \and
Harald Sack\inst{1,2}\orcidID{0000-0001-7069-9804}
}
\authorrunning{Bruns et al.}
\titlerunning{NFDIcore 2.0}
%
\institute{FIZ Karlsruhe – Leibniz Institute for Information Infrastructure, Hermann-von-Helmholtz-Platz 1,
76344 Eggenstein-Leopoldshafen, Germany
\email{firstname.lastname@fiz-karlsruhe.de}\\
\and
Karlsruhe Institute of Technology (AIFB), Kaiserstr. 89, 76133 Karlsruhe, Germany
}
\maketitle              
\begin{abstract} 
This paper presents NFDIcore 2.0, an ontology compliant with the Basic Formal Ontology (BFO) designed to represent the diverse research communities of the National Research Data Infrastructure (NFDI) in Germany. NFDIcore ensures the interoperability across various research disciplines, thereby facilitating cross-domain research. Each domain's individual requirements are addressed through specific ontology modules. This paper discusses lessons learned during the ontology development and mapping process, supported by practical validation through use cases in diverse research domains. 
The originality of NFDIcore lies in its adherence to BFO, the use of SWRL rules for efficient knowledge discovery, and its modular, extensible design tailored to meet the needs of heterogeneous research domains. 

\end{abstract}
\section{Introduction} 
Knowledge Graph supported research data infrastructures are currently being developed in domain specific\footnote{\url{https://netwerkdigitaalerfgoed.nl/en/}}, national~\cite{hyvonen2020linked}, and international\footnote{\url{https://www.clarin.eu/}}\footnote{\url{https://sshopencloud.eu/}} efforts to facilitate advanced data discovery, scientific collaboration, and innovation. The National Research Data Infrastructure (NFDI)\footnote{\url{https://www.nfdi.de}} is a German national initiative with the goal to provide an organized, standardized, and sustainable research data infrastructure for diverse research domains, which are covered by respective NFDI consortia, represented by universities, research institutes, and infrastructural organizations all over the country. For example, NFDI4Culture\footnote{\url{https://nfdi4culture.de/}} is a consortium for research data on material and immaterial cultural heritage \cite{altenhoner_2020,bruns2024damalos,tietz2023damalos}, NFDI-MatWerk\footnote{\url{https://nfdi-matwerk.de/}} is providing infrastructure for Materials Science Engineering~\cite{eberl_2021}, NFDI4DataScience\footnote{\url{https://www.nfdi4datascience.de/}} in the domains of Data Science \& Artificial Intelligence, MaRDI\footnote{\url{https://www.mardi4nfdi.de/}} for mathematical research data, NFDI4Memory\footnote{\url{https://4memory.de/}} for historical research~\cite{paulmann_2022}, etc.

The NFDI consortia share similar overarching goals and concepts, including their structure, governance, persons, institutions, areas of expertise, data repositories, devices, services, and more~\cite{sack2023cordi}. Additionally, each consortium has individual needs for research data discovery and reuse, and requires specific standards and tailored solutions to address their unique challenges. Ensuring the interoperability of the research (meta) data among these NFDI consortia has significant benefits and allows for knowledge discovery across research domains, such as: \textit{"Which materials contribute to the preservation and restoration of cultural heritage objects?", "What mathematical formulas were employed in the production and distribution of goods in ancient and medieval economies?", "What data science techniques are applied to the study of archaeological artifacts to identify ancient trade routes?"}.

To achieve a high level of interoperability, and hence to be able to answer these questions, a data model is needed that represents the overarching concepts of the domains on the one hand and meets the specific requirements of the individual communities on the other. To address this, the following research questions (RQ) were developed: 

\begin{itemize}[leftmargin=4em]
    \item [RQ 1:] How can the overarching concepts of heterogeneous research domains be modeled to ensure a comprehensive representation and integration of all disciplines and their research (meta)data?
    \item [RQ 2:] How can the various standards used across research domains be effectively mapped and integrated to ensure interoperability across diverse research disciplines?
    \item [RQ 3:] How can the NFDIcore ontology be extended and tailored to address the specific needs of research consortia from heterogeneous domains?
    \item [RQ 4:] What impact does the integration of complex relations, processes and roles in NFDIcore have on the usability, expressivity and interoperability of the ontology?
    \item [RQ 5:] How can the representation and reasoning of complex roles, processes, and relationships in NFDIcore be improved?
\end{itemize}

This paper contributes NFDIcore, the ontology compliant to the Basic Formal Ontology (BFO) \cite{bfo2015} that represents NFDI consortia from various research disciplines to ensure interoperability between them and hence, facilitate cross-domain research in a modular approach. The individual requirements of each respective research domain or consortium is addressed by means of ontology extensions. This paper furthermore contributes a discussion of lessons learned during the ontology development and mapping process and describes the practical evaluation through real-world use cases. The originality of the presented approach lies in its combination of compliance with the well-established BFO as an upper ontology, the introduction of short-cuts by means of SWRL rules\footnote{\url{https://www.w3.org/submissions/SWRL/}}, and a modular and extensible design tailored to various research domains. 

The paper is structured as follows: Section~\ref{sec:relatedwork} discusses related work. Section~\ref{sec:nfdicore} details the ontology design and mapping workflow and discusses the lessons learned in the process.  Section~\ref{sec:usecase} contains proof-of-concepts by means of real-world use cases. Section~\ref{sec:conclusion} concludes this paper.
\section{Related Work}
\label{sec:relatedwork}
\sloppy
Ontologies have been used to represent and interconnect FAIR research (meta)data and to facilitate data exploration across platforms, data sets and repositories. The VIVO ontology \cite{corson2012} represents researchers and the full context in which they are working, including their outputs, interests, accomplishments, and associated institutions. The ontology was created for the VIVO software and integrates a number of existing standards and ontologies, such as BFO, Dublin Core\footnote{\url{https://www.dublincore.org/resources/glossary/ontology/}}, the Event ontology\footnote{\url{http://motools.sourceforge.net/event/event.html}} and FOAF\footnote{\url{http://www.foaf-project.org/}}. VIVO represents researchers and their work extensively, but also in a rather complex and fine grained manner. In contrast, this paper presents an approach in which heterogeneous research domains are integrated in a BFO-compliant mid-level ontology while domain specific requirements are modeled by means of modular extensions. 
The DCAT vocabulary\footnote{\url{https://www.w3.org/TR/vocab-dcat-3/}} enables publishers to describe datasets and data services in a catalog on the Web. The goal is to ensure discoverability, enable a decentralized approach to publishing data catalogs and providing the possibility for federated search for datasets across catalogs in multiple sites. The Core Data Set for Research (KDSF)\footnote{\url{https://kerndatensatz-forschung.de/}} represents research information for the German academic system with the goal to harmonize and standardize the research reporting system at universities and research institutions. However, KDSF is too specific in representing the German academic system, which is not in the scope of the goals of this work. Schema.org is a community-driven standard to structure data on the Web, including research data, and facilitates the integration with Web technologies, making it easier to discover and use research data. However, utilizing schema.org alone would not suffice for the requirements of this work due to its limited semantic expressivity and incompatibilities with the ontologies and standards of some domains (e.g., Material Science).

As a mid-level ontology, NFDIcore facilitates the interoperability across the NFDI consortia to represent metadata about research resources such as individuals, organizations, projects, data portals, services, etc. NFDIcore establishes mappings to standards across domains, including the Basic Formal Ontology (BFO) and schema.org, which is crucial for advancing knowledge representation, data exchange, and collaboration across the diverse domains.

\section{The NFDIcore Ontology}
\label{sec:nfdicore}
This section reflects the lessons learned from the development and implementation of NFDIcore~1.0, provides a comprehensive overview of the NFDIcore~2.0 ontology, detailing the methodology employed, the design principles, and the key components that make up the ontology. It also addresses how NFDIcore~2.0 balances expressivity with usability, ensuring that complex relationships can be accurately represented while maintaining ease of integration and accessibility for users across various research domains. 
\subsection{Lessons Learned}
The development and evolution of the NFDIcore ontology emerged from a series of critical observations and challenges encountered throughout the NFDI4Culture project and subsequent collaborative efforts with other NFDI consortia. The following points outline the key lessons learned, which have shaped the motivation and design of NFDIcore~2.0: 

\noindent\textbf{Initial Scope of the Ontology.} The development of the first version of the NFDIcore ontology was based solely on the data at hand, primarily focusing on contribution descriptions and metadata about resources such as data portals, datasets, and services contributed to the NFDI4Culture project, which was among the first NFDI projects to start. The ontology was developed following a bottom-up approach, which facilitated a lightweight design. However, the lack of data and limited collaboration in the initial project phase resulted in an ontology that did not account for the broader needs of the culture community or potential future requirements from other NFDI project communities, like materials science or data science. 

\noindent\textbf{User Engagement and Feedback Collection.} There was a lack of early and continuous engagement with the community to understand their specific needs and expectations. This resulted in a disconnect between the ontology's design and the community's expectations, and led to resistance in using the ontology and changing established practices in favour of usability and interoperability. 

\noindent\textbf{Foundational vs Lightweight.} The prioritization of a lightweight representation in the initial versions of the ontology hindered an effective integration for domains requiring rigorous representations of complex relationships and processes. For example, NFDICore~1.0 was able to only capture the basic fact of dataset creation by an agent (\texttt{<:Dataset :creator :Agent>}), while, for researchers in the domain of materials science it is essential to describe not only the creation itself but also the inputs, methodologies, experimental conditions, and outputs of the creation process. 

\noindent\textbf{Ambiguous Representation.} Lightweight representations often lead to ambiguities. For example, many schema.org classes belong to multiple hierarchies, making them challenging to interpret and organize -- \textit{schema:CreativeWork} covers entities such as data sets, books, movies, and artworks, which are typically tangible outputs of creative projects. On the other hand, \textit{schema:Service} encompasses intangible offerings like reservations, ticketing services, and subscriptions. A "knowledge graph" could be seen as both a creative work due to its creation process, and a service, since it provides a structured way to access and navigate knowledge. Defining it as a subclass of both might effect reasoning and inference. In contrast, more structured and foundational ontologies provide explicit definitions, relationships, and constraints to ensure clarity and precision in data modeling across domains. \\
Overall, while practical at the early stage of the project, the lightweight ontology design led to challenges in data integration and representation across domains. Initially, it was unclear what the coverage should be, and the need for a foundational and universal framework was not recognized. Therefore, starting with lightweight representations based on the available data was considered appropriate and allowed for the definition of the central concepts of NFDIcore based on the collected feedback. In the next step, to adapt to the requirements of the community, NFDIcore~2.0 was created on the basis of the NFDIcore~1.0. On the one hand, NFDIcore addresses the lessons learned and provides ways for robust unambiguous representations via mapping to BFO -- a foundational ontology that focuses on covering universal concepts across domains. On the other hand, NFDIcore~2.0 provides a more flexible framework via integrating essential schema.org classes, and providing shortcuts for complex relations that allow for more lightweight representation and querying of complex relations when this is required. This ensures that both detailed and simplified knowledge representations coexist, enhancing usability and integration. In the next section, the NFDIcore design methodology and central components will be introduced. 

\subsection{Methodology and Design}
The NFDI programme is overall funding 26 consortia from all scientific disciplines. The creation of NFDIcore began with NFDI4Culture, a consortium from the first of three funding rounds. In the initial phase of NFDI4Culture, the foundation of the ontology was established based on the contribution descriptions -- metadata about research resources provided by data contributors during the proposal phase. The development of the NFDIcore ontology is guided by a bottom-up, iterative approach, adhering to user-centered design principles. This was enriched through collaboration with culture domain experts, resulting in the creation of user stories\footnote{\url{https://nfdi4culture.de/resources/user-stories.html}}. These user stories were formulated through a series of meetings and practical workshops, ensuring that the ontology was shaped by real-world needs and scenarios.

Following the methodology established by NFDI4Culture, further consortia also contributed their user stories and requirements. For instance, the NFDI-MatWerk community developed infrastructure use cases\footnote{\url{https://nfdi-matwerk.de/infrastructure-use-cases}}, NFDI4DS crafted detailed personas\footnote{\url{https://www.nfdi4datascience.de/community/requirements-elicitation/personas/}}, and NFDI4Memory identified specific problem stories\footnote{\url{https://4memory.de/problem-stories-overview/}}. These inputs from different NFDI consortia highlighted various community-specific challenges, and pointed out requests shared across the consortia. By systematically examining the common challenges and requirements (see e.g. \cite{tietz2023damalos}), the central components of the ontology were identified.

\begin{figure}[t]
    \centering
    \includegraphics[width=0.9\textwidth]{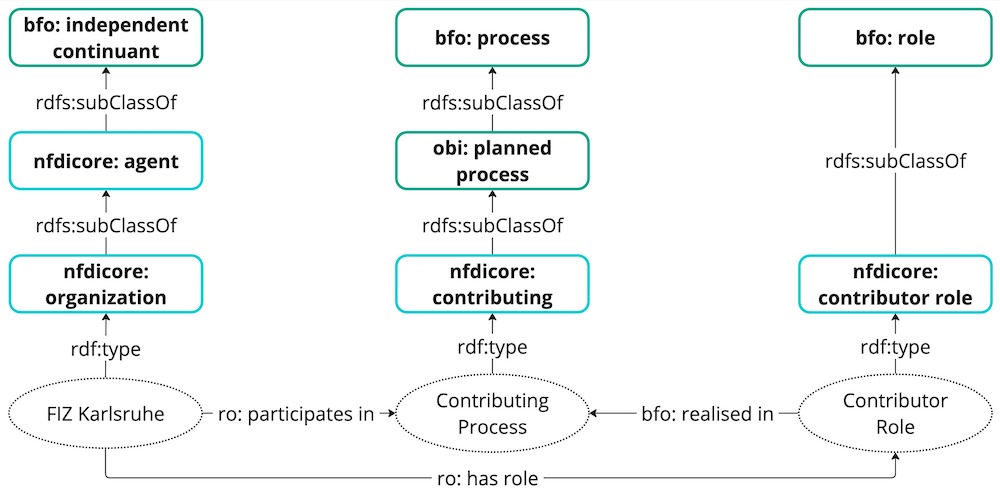}
    \caption{Representing independent continuants and their context roles in NFDIcore}
    \label{ic+roles}
\end{figure}

Following the best practices of ontology development~\cite{rudnicki2016best} and the requirements of versatile NFDI communities, NFDIcore aims at reusing the existing ontologies and extending them according to the requests of the community. Addressing the lessons learned, e.g. the need for universal foundational representations, the objectives of NFDI, and the principles the ontology is committed to~\cite{otte2022bfo}, BFO - Basic Formal Ontology~\cite{bfo2015} has been selected to serve as a top level ontology of the NFDIcore~2.0. In addition, parts of further ontologies, as e.g., schema.org and DCAT, are used to ensure smooth integration. \\
 \\
\textbf{Continuants in NFDIcore}\\
In BFO, a distinction is made between \textit{continuants}, which are fully present at any time they exist, and \textit{occurents}, which unfold over time or have temporal elements. A special case of continuants are \textit{independent continuants}, whose existence does not depend on any other entities. In NFDI, \textit{agents} such as organizations and persons play active roles in various processes. Thus, by defining  \textit{nfdicore:agent} as a subclass of \textit{bfo:independent continuant}, agents are recognized as entities capable of action without being dependent on other entities for their existence. \\
\textit{Independent continuants} in BFO are further distinguished into two subtypes: \textit{material entities}, which are spatially extended into three dimensions, and \textit{immaterial entities}, which refer to specific spatial regions or locations of independent continuants. \textit{nfdicore:Collection} is a material entity encompassing aggregates of tangible items such as a collection of arts displayed at the museum, a collection of historical books, a collections of different materials (metals, polymers, ceramics), etc. Following the requirements for more granular representation of locations, immaterial entity \textit{bfo:site} is further extended with \textit{nfdicore:Place, nfdicore:Country, nfdicore:City, nfdicore:FederalState}. \\
One of the distinguishing characteristics of the \textit{independent continuants} from other continuants is that they can serve as bearers of roles, meaning they can have roles that are not intrinsic to their nature but are relevant in specific contexts. For example, an organization can be a member of a consortium (\textit{nfdicore:ConsortiumMemberRole}), and act as a funder of a project (\textit{nfdicore:FundingOrganizationRole}). To specify the context in which the role is realized, properties \textit{bfo:realized in}, \textit{ro:has role}\footnote{ro: \url{http://purl.obolibrary.org/obo/ro.owl}} and \textit{ro:participates in} are used (see example in Figure~\ref{ic+roles}\footnote{obi: \url{http://purl.obolibrary.org/obo/obi.owl}}). 
 \\
 \\
\textbf{Research Artifacts}\\
The central components across all consortia of NFDI are research resources that encompass a wide range of digital creative works, including datasets, collections, and metadata, as well as offered products and services such as data portals, data curation, and data digitization. Within the BFO ontology network, the Information Artifact Ontology (IAO)\footnote{\url{https://github.com/information-artifact-ontology/IAO/}} was created to serve as a domain‐neutral resource for the representation of types of	\textit{information	content	entities} (ICEs) such as documents, data‐bases, and	digital	images. ICEs are defined as \textit{generically dependent continuants}, meaning their existence depends on the existence of another entity. In NFDIcore, the central class \textit{nfdicore:Resource} is categorized as an \textit{iao: information content entity}\footnote{iao: \url{http://purl.obolibrary.org/obo/iao.owl}} derived from a \textit{bfo: material entity}.  For example, a dataset of digital images might denote a tangible art collection in a museum. In \textit{nfdicore:Resource} a distinction is made between \textit{nfdicore:CreativeWork} and \textit{nfdicore:Service}. For the purposes of smooth integration and to meet community requirements, these classes provide a direct mapping to schema.org. However, unlike \textit{schema:CreativeWork}, \textit{nfdicore:CreativeWork} is specialized to cover only digital creative works.
\begin{quote}
{\small \tt
\begin{tabbing}
nfdicore:CreativeWork \=$\equiv$ \=(iao:information content entity $\cap$ \\ \> \> \, schema:CreativeWork)
\end{tabbing}
 }
 \end{quote}
\noindent This specialization, through the mapping to \textit{schema:CreativeWork} and \textit{information content entity}, excludes material creative works such as real paintings, architectural objects, and other physical artifacts, which fall outside the scope of \textit{nfdicore:CreativeWork}. Similarly, \textit{nfdicore:Service} is equivalent to the intersection of \textit{iao: information content entity} and \textit{schema:Service} to exclude services, which are out of the scope of the NFDIcore.

\begin{figure}[t]
    \centering
    \includegraphics[width=1\textwidth]{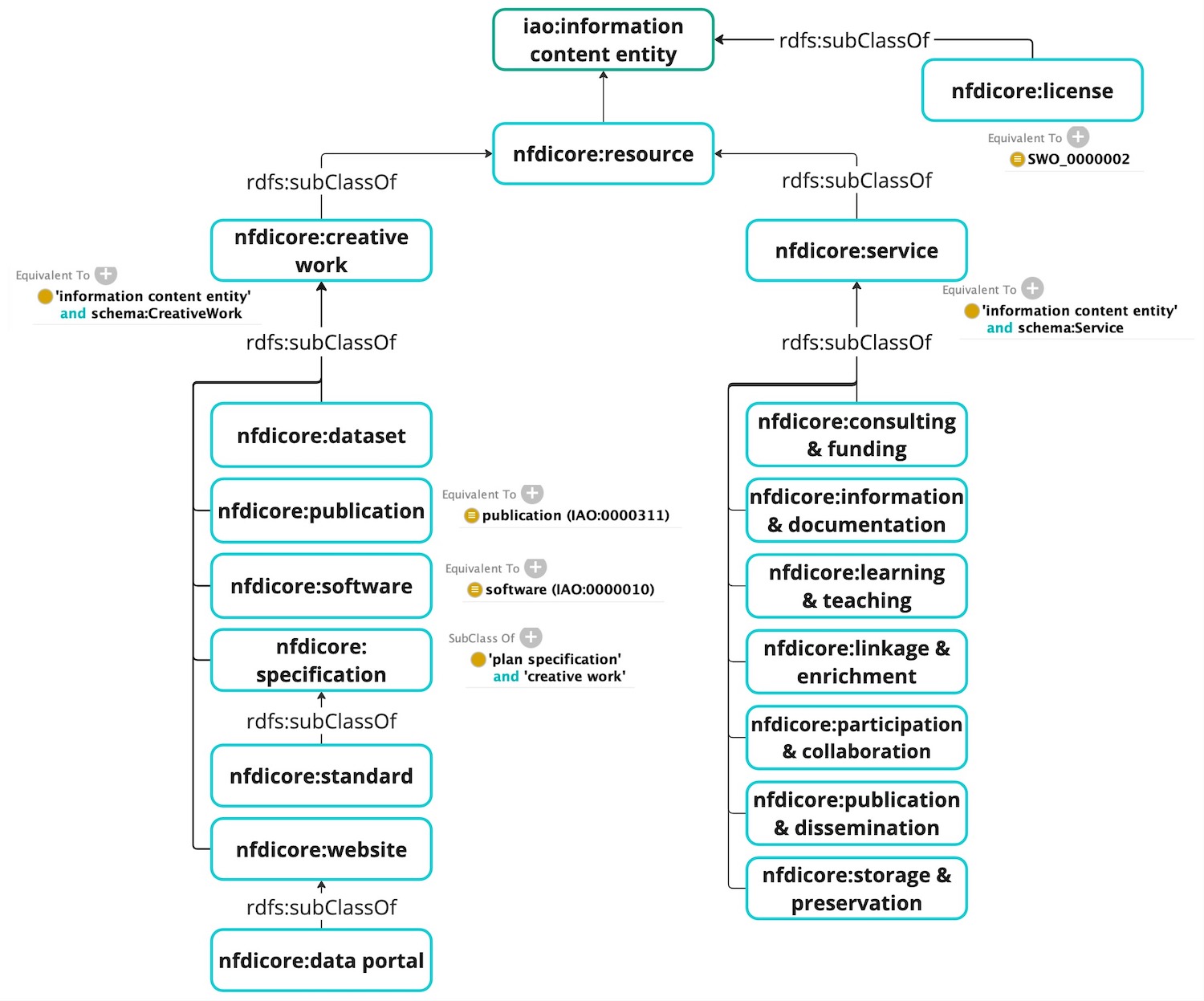}
    \caption{Information Content Entities in NFDIcore.}
    \label{ices}
\end{figure}

\begin{figure}[t]
    \centering
    \includegraphics[width=1\textwidth]{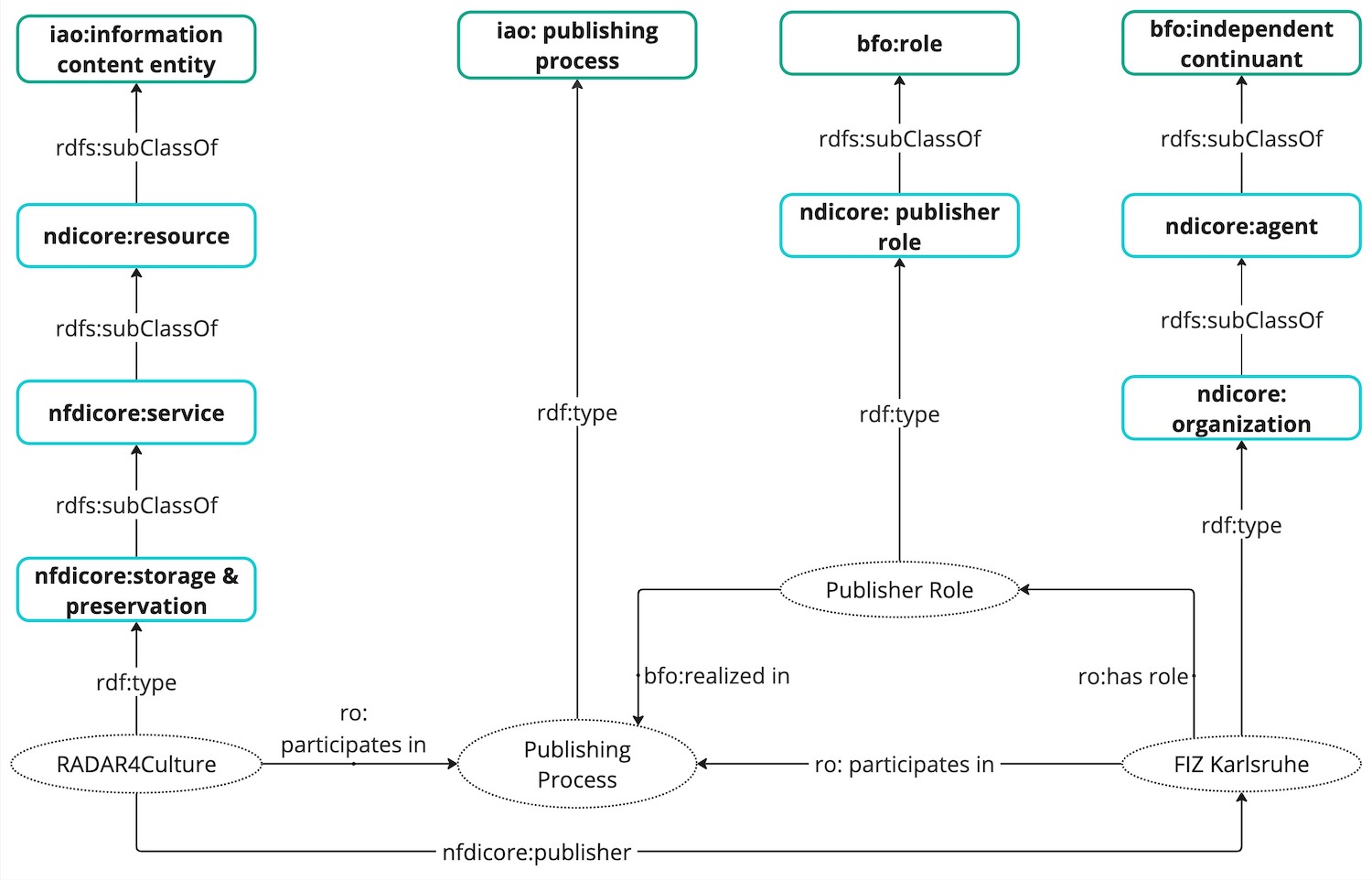}
    \caption{Example of representing NFDIcore entities and their relations}
    \label{process}
\end{figure}

\medskip
\centerline{\small {\texttt{nfdicore:Service} $\equiv$ (\texttt{iao:information content entity }$\cap$ \texttt{schema:Service})}}
\medskip
\noindent Based on the community input, \textit{nfdicore:CreativeWork} and \textit{nfdicore:Service} are categorized into more fine grained resources such as \textit{data sets}, \textit{data portals}, \textit{software}, which are linked to corresponding classes in BFO-compliant ontologies (e.g., IAO, SWO\footnote{https://github.com/allysonlister/swo}), wherever feasible (see the full hierarchy in Figure~\ref{ices}). However, these categories are not exhaustive. Should a consortium require additional domain-specific classes to represent their resources, they are able to add them in their domain extensions of the ontology.\\
\begin{figure}[t]
    \centering
    \includegraphics[width=1\textwidth]{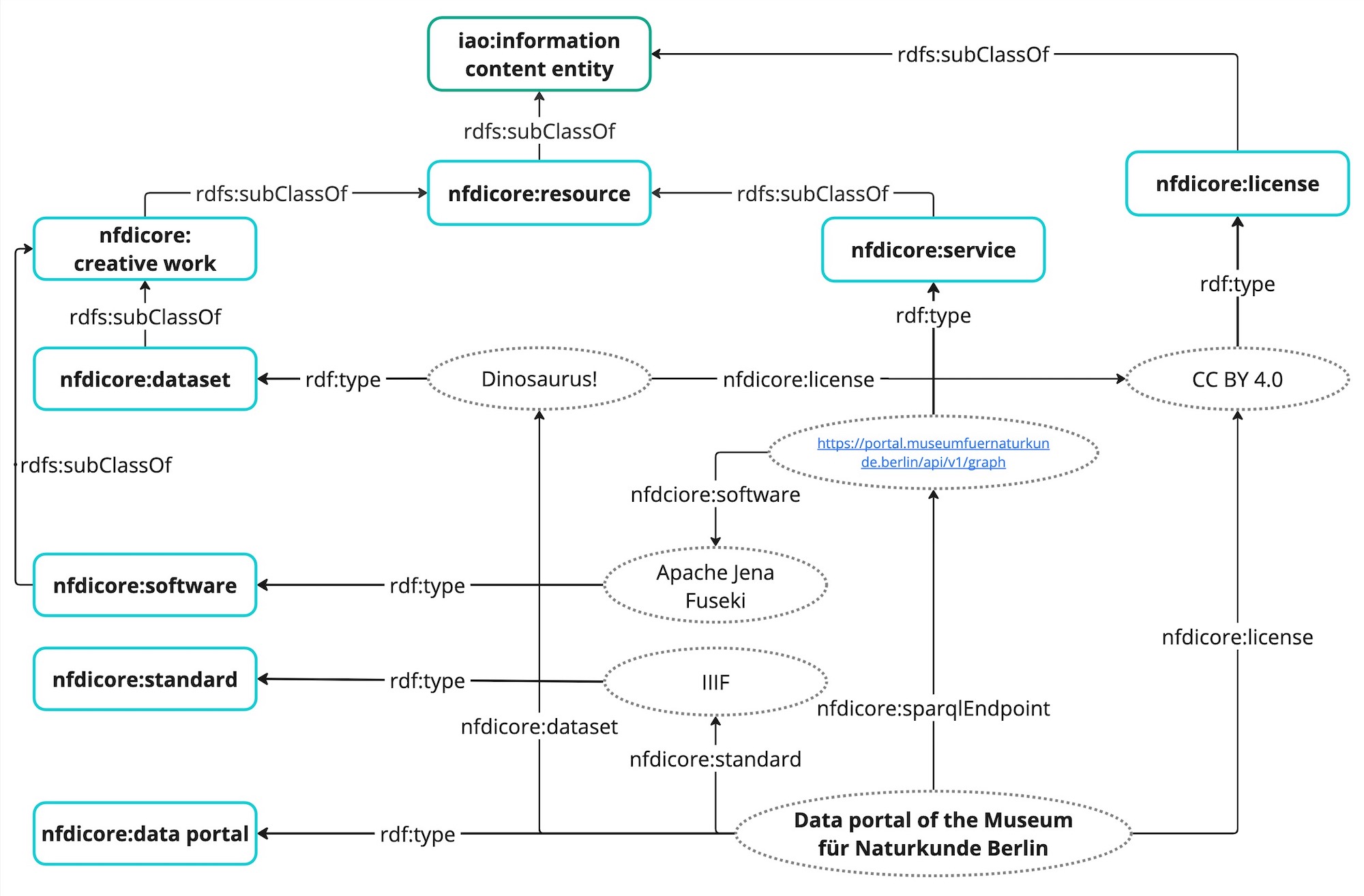}
    \caption{Example of representing resources and their relations in NFDIcore}
    \label{ices2}
\end{figure}
Information content entities in IAO stand in an \textit{aboutness} (\textit{iao:is about}) relation to other entities. While this provides a basic linkage between ICEs and other entities, the NFDI community requires more nuanced relationships to accurately capture the specifics of how resources are related to each other. Thus, subproperties of the \textit{iao:is about}  such as \textit{nfdicore:license}, \textit{nfdicore:software}, \textit{nfdicore:sparqlEndpoint} to model more detailed relationships between resources are introduced (see example in Figure~\ref{ices2}). Apart from aboutness, resources within NFDIcore can also be described using specific \textit{qualities}. For instance, resources can be associated with \textit{academic disciplines} they belong to or a degree of \textit{semantic expressivity} the resources hold, allowing for a more nuanced understanding of their context and relevance. For this, the property \textit{ro:has quality} is extended with more finegranular nfdicore properties. \\
 \\
\textbf{Occurents in NFDIcore}\\
In BFO, a \textit{process}  encompasses an \textit{occurent} that occurs deliberately rather than naturally. A \textit{planned process} is a special type of \textit{bfo:process}, which was originally suggested by Ontology for Biomedical Investigations (OBI)\footnote{\url{https://obi-ontology.org/}}, designed to encompass processes that are conducted according to a predefined plan. In NFDIcore, \textit{obi:planned process} and its subclasses, e.g. \textit{iao:publishing process}, \textit{nfdicore:Project, \textit{nfdicore:Contributing}}, is used to establish complex connections between NFDIcore entities, as depicted in Figure~\ref{process}. \textit{nfdcore:Event} is defined as subclass of \textit{bfo:occurent} and represents entities that occur or unfold over time, typically characterized by specific happenings or occurrences with temporal properties. Events in NFDI encompass a wide range of occurrences, including but not limited to conferences, wars, strikes, celebrations, accidents, workshops, meetings, and other activities.

\subsection{Usability and Shortcuts}
By leveraging processes and roles, NFDIcore~2.0 enables a detailed representation of how different entities interact and relate to one another over time. This is beneficial for documenting research workflows, collaborations, and contributions within research projects, ensuring that the multifaceted nature of research activities is captured accurately and precisely. This enhances the ontology's level of expressivity and supports a more comprehensive data integration and analysis. However, to address the community's requirements for easier integration, especially for those who do not use complex relations in their representations and prefer straightforward lightweight approaches, and to ensure that NFDIcore~2.0 remains compatible with NFDIcore~1.0, where no complex relations were used, NFDIcore~2.0 offers a set of SWRL rule-based shortcuts. This approach benefits users by simplifying the ontology, making it more accessible and easier to implement, thus encouraging broader adoption. Figure~\ref{process},
 depicts the visualization of the shortcut relation \textit{nfdicore:publisher}. The respective rule in SWRL\footnote{The prefixes are omitted for readability reasons.}:  
 \begin{quote}
{\small \texttt{Agent(?a) $\wedge$ Resource(?r) $\wedge$ PublisherRole(?pr) $\wedge$ publishing process(?p) $\wedge$ participates in(?a, ?p) $\wedge$ participates in(?r, ?p) $\wedge$ has role(?a, ?pr) $\wedge$ realized in(?pr, ?p) $\rightarrow$ publisher(?r, ?a)} 
}
 \end{quote}
The execution of this rule creates a new relationship \textit{publisher}, which directly connects the \textit{agent} with the \textit{resource}, indicating that the \textit{agent} is the \textit{publisher} of the \textit{resource}.

This solution facilitates the smooth integration of straightforward relationships while maintaining the ontology's alignment with BFO principles. By adopting this approach, NFDIcore~2.0 provides the flexibility to use both complex and simple relations, while the shortcuts ensure compatibility and simplify querying, making it easier to identify connections regardless of the complexity of the relationships used.
Further SWRL rules integrated in NFDIcore are documented on the Web\footnote{\url{https://ise-fizkarlsruhe.github.io/nfdicore/\#swrlrules}}.

\subsection{Modules of NFDIcore}
\begin{figure}[t]
    \centering
    \includegraphics[width=1.0\textwidth]{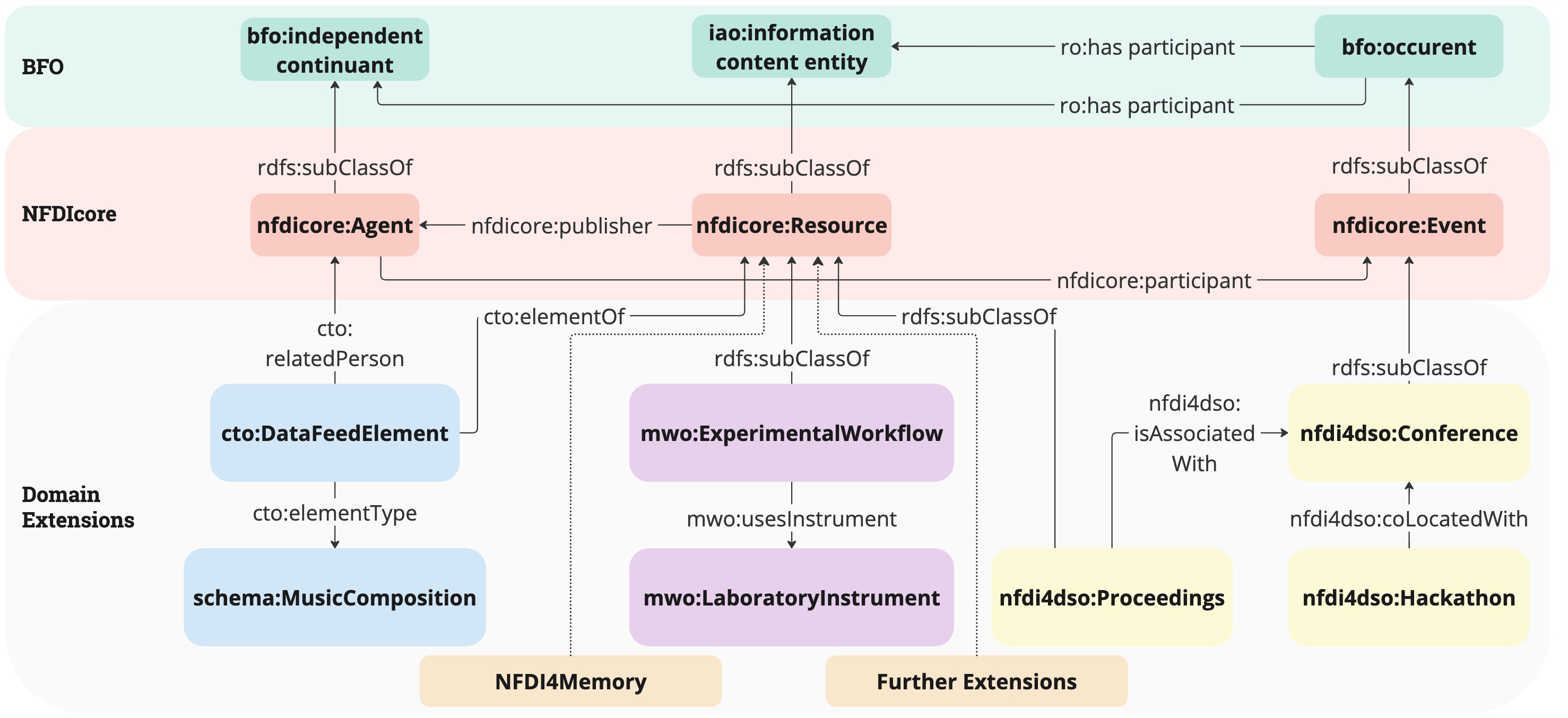}
    \caption{A visualization of the layers and modules of NFDIcore.}
    \label{fig:modules}
\end{figure}
While NFDIcore~2.0 represents the overarching concepts of the NFDI consortia, the specific needs for research (meta)data representation in each consortium are addressed by means of ontology extensions, or modules~\cite{sack2023cordi}. So far, three modules (cf. Figure~\ref{fig:modules}) have been released and connected with NFDIcore, with further modules currently in preparation.
The NFDI4Culture Ontology (CTO)\footnote{\url{https://gitlab.rlp.net/adwmainz/nfdi4culture/knowledge-graph/culture-ontology}} is a module designed to represent and categorize resources within the NFDI4Culture domain, which encompasses five academic disciplines: Architecture, Musicology, Art History, Media Science, and the Performing Arts. CTO defines classes and relations that address domain-specific research questions, connect diverse cultural entities, and facilitate the efficient organization, retrieval, and analysis of cultural data~\cite{bruns2024damalos}. The MatWerk ontology (MWO)\footnote{\url{https://nfdi-matwerk.pages.rwth-aachen.de/ta-oms/mwo/docs/index.html}} serves as the backbone for the materials science and engineering knowledge graph (MSE-KG). The graph focuses on the relevant community structure, infrastructure like software, workflows etc., and data repositories, databases etc. The NFDI4DS ontology (NFDI4DSO)\footnote{\url{https://github.com/ISE-FIZKarlsruhe/NFDI4DS-Ontology/}} represents metadata on resources in the data science domain and content-related index data, as e.g., metadata for training datasets and machine learning models. The NFDI4Memory module is currently in development and in a first stage captures information about research structures within the consortium, including archives, libraries, museums, and other university institutions~\cite{ondraszek2024}.
This modular architecture allows each consortium to integrate their data into individual KGs and research data platforms (e.g. the Culture Information Portal\footnote{\url{https://nfdi4culture.de/}}). At the same time, this presented approach enhances the interoperability between consortia and allows for cross-domain queries, e.g.  "Which materials contribute to the preservation and restoration of cultural heritage objects?", "What mathematical formulas were employed in the production and distribution of goods in ancient and medieval economies?".

\section{NFDIcore Evaluation} 
\label{sec:usecase}

\begin{figure}[t]
    \centering
    \includegraphics[width=1\textwidth]{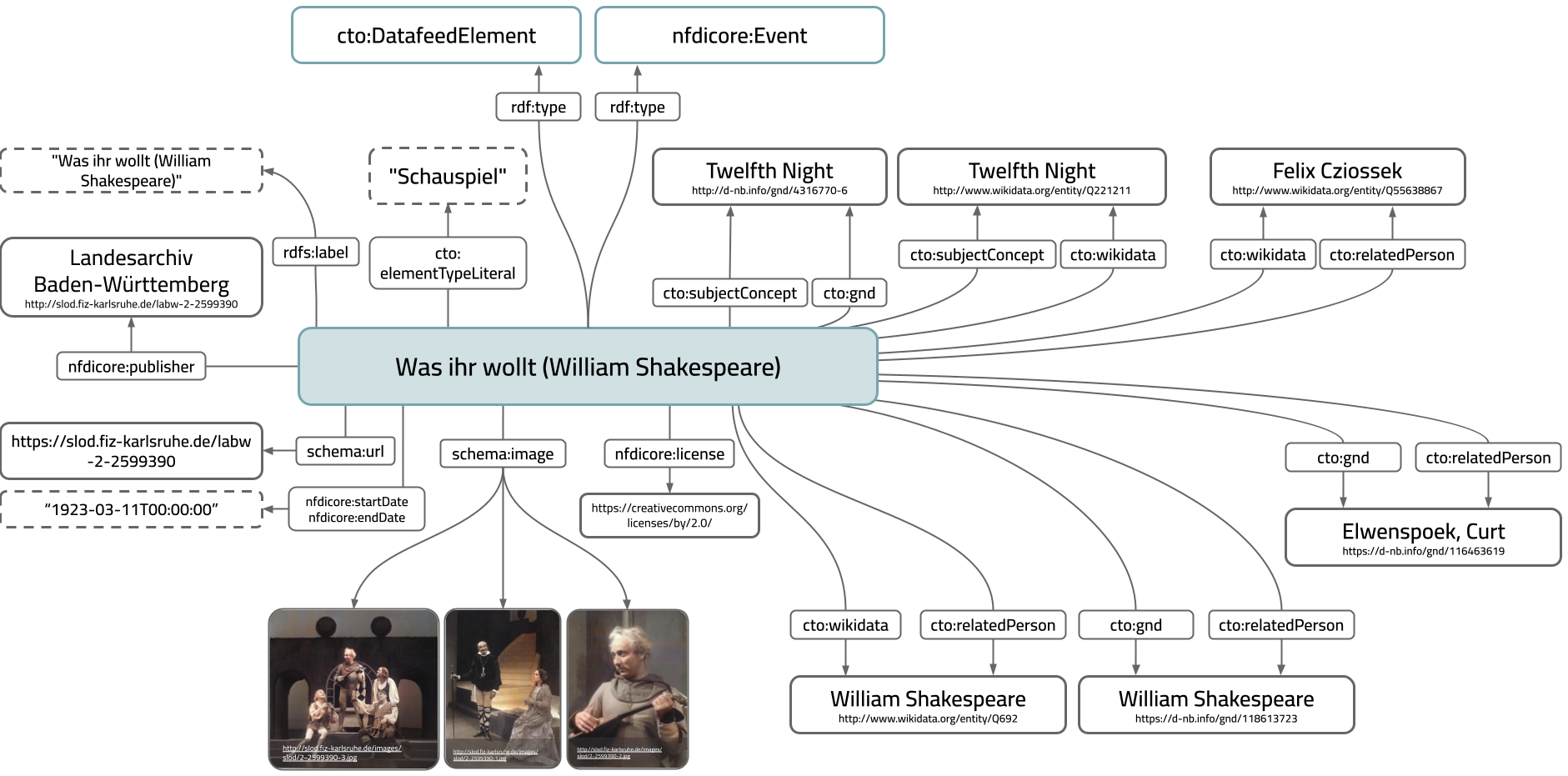}
    \caption{An example of representing a performing arts event "Was Ihr Wollt" with CTO.}
    \label{cto-pa}
\end{figure}

The starting point of evaluating NFDIcore was the collection of detailed user stories that describe specific tasks, challenges, and requirements related to research and data management. The use cases were provided by domain experts across four NFDI consortia: NFDI-MatWerk, NFDI4Culture, NFDI4DataScience, and NFDI4Memory. The user stories were analyzed to develop a set of concrete competency questions (CQs), which were to be expressed in SPARQL queries for the ontology evaluation. During the preparation process, each use case was reviewed to identify the core requirements and objectives. \\
The collected use cases often address specific domain questions, reflecting the unique needs and terminologies of each research area. However, similarities are identified across every consortium. For example, domain experts across all disciplines ask for services: in NFDI4Culture, services for creating digital inventories of art collections; in NFDI-MatWerk, services for storing and evaluating raw mechanical test data; and so forth. Thus, for evaluating NFDIcore, this specific domain information was omitted to create a set of more generalized CQs fitting across consortia. For example, from a use case where a musicologist searches to learn more about music notation standards such as MEI by contacting a responsible person, the essential requirements were translated into more generalized CQs: \textit{What standards are there for specific resource types? Who is a contact point for the standards?} This approach ensures that the ontology remains flexible and applicable across diverse research domains without being restricted by the domain-specific peculiarities of each consortium.\\
Overall, there are 98 CQs collected across consortia. The analysis of the CQs reveals several recurring concepts that are of main interest across the different consortia:
\begin{itemize}
    \item  \textbf{Services}, e.g. What services are available for specific data task, \textit{e.g. keeping different data versions accessible}? (NFDI4Culture) What specific services, \textit{e.g. in digitization}, are used in a specific academic discipline, \textit{e.g. data science}? (NFDI4DataScience)
\item \textbf{Standards and Specifications}, e.g. What standards and specifications are used in a certain process, \textit{e.g. data analysis}? (NFDI-MatWerk) What are the best practices to publish a resource, \textit{e.g. archival data}, using a certain license? (NFDI4Memory)
\item \textbf{Processes}, e.g. What resources and events are related to specific processes, \textit{e.g. demonstration and teaching}? (NFDI-MatWerk) What documenting resources are related to a specific event (task)? Who is a contact point of the documentation process? (NFDI4Culture)
\item \textbf{Events}, e.g. List the events in the community, their description and their dates. (NFDI4DataScience) What services and events are related to specific processes, \textit{ e.g. structured filing and the handling of standard data}? (NFDI4Culture) 
\item \textbf{Contact Points}, e.g. Who is a contact point of a specific resource, \textit{e.g. experimental dataset}? What is their expertise? (NFDI-MatWerk) Who has a certain area of expertise, \textit{e.g. IT} and is involved in a certain service, \textit{e.g. training and education}? (NFDI4Memory) 
\end{itemize}
The results of the evaluation have shown that the vast majority of the CQs could be directly translated to SPARQL queries\footnote{\url{https://github.com/ISE-FIZKarlsruhe/nfdicore/tree/main?tab=readme-ov-file\#nfdicore-competency-questions-cqs-and-sparql-queries}}. Due to the flexible ontology design, both the complex role information and lightweight representations can be queried effectively. For example, when retrieving learning and teaching services and their contact points:
\begin{quote}
\texttt{SELECT ?contactPoint ?service \\
WHERE \{ \\
  ?contactPoint bfo:RO\_0000087 ?role ;\\
  bfo:RO\_0000056 ?process . \\
  ?role bfo:BFO\_0000054 ?process .\\
  ?role rdf:type nfdicore:ContactPointRole .\\
  ?service bfo:RO\_0000056 ?process ;\\
  rdf:type nfdicore:Learning\&Teaching .\}
  }
  \end{quote}
 The same result can be achieved by a more user-friendly and efficient query,  utilizing the inferred relationships established through the SWRL rule.
\begin{quote}
\texttt{SELECT ?service ?contactPoint \\
WHERE \{
  ?service rdf:type nfdicore:Learning\&Teaching ;
  nfdicore:contactPoint ?contactPoint .\}
  }
\end{quote}
This flexible design enhances NFDIcore's ability to represent and reason about complex roles, processes, and relationships. This results in improved performance, usability, and collaborative potential, making the ontology fit for diverse research and data management needs across consortia.\\
Several CQs could not be translated to SPARQL queries since they require domain-specific knowledge, which is outside the scope of NFDIcore, e.g. CQs about specific elements in a data resource. For example, retrieving all data set elements related to Shakespeare's "Twelfth Night". Since the representation of elements in resources is highly domain specific, such CQs are to be addressed by the extension modules. For example, Figure~\ref{cto-pa} visualizes the representation of the performing event "Was Ihr Wollt" with CTO.\\ 
Three CQs could not be translated into SPARQL queries, highlighting the need for further discussion regarding the NFDIcore. In contrast to other consortia, NFDI4DataScience treats \textit{processes} as \textit{resources} within the framework of NFDIcore. For instance, to answer the CQ: “What standards are there for a specific process, \textit{e.g. sharing data}?”, the ontology has to be adapted, since currently, NFDIcore addresses standards as resources about (only) other resources via subproperties of \textit{iao:is about}. While, similar to other resources, a standard can \textit{participate in} a process, this does not imply that this standard is specifically for executing the process. This gap presents the need for NFDIcore to explicitly relate standards to processes and hence, to reflect their role in the processes. However, this has to be addressed in a way that avoids inconsistencies with BFO, since in BFO a role can only be borne by \textit{independent continuants}. As this issue has currently been spotted in one consortium, it remains open. Further discussions are essential as additional consortia adopt the ontology, determining whether these challenges should be addressed within NFDIcore itself or through domain-specific extensions like DSO.\\
The evaluation of NFDIcore across four different consortia has demonstrated its effectiveness, adaptability and role as a foundational framework for representing collaborative research and data management across diverse consortia and academic disciplines, while also highlighting areas for future development and enhancement.

\section{Discussion and Conclusions}
\label{sec:conclusion}
The development and evaluation of NFDIcore 2.0 provide valuable insights into representing and integrating 
heterogeneous research domains. NFDIcore~2.0 addresses these complexities by exploiting the foundational ontology BFO, ensuring a universal representation of core concepts (RQ1). By incorporating established standards and external ontologies like schema.org and DCTERMS, NFDIcore~2.0 facilitates effective data integration across research domains (RQ2). Moreover, NFDIcore~2.0 adopts a modular approach to ontology design, allowing for the development of domain-specific ontology modules. This enables each consortium to extend the core ontology to meet the specific needs pf their domain (RQ3). Mapping NFDIcore to BFO and BFO-compliant ontologies enriches NFDIcore~2.0 with complex relations, processes, and roles. It enhances the ontology's expressivity and usability through detailed representation of entities and their interactions, supports comprehensive data integration and analysis, reducing ontology's ambiguity. However, such complexity can pose challenges in ensuring consistency and interoperability across domains, when aligning it with external standards or ontologies existing in different disciplines. Additionally, while a detailed model provides richer semantics, their use may also become challenging for less experienced users (RQ4). The incorporation of shortcut relations aims at addressing this challenge by allowing for integrating data that lack complex representations. The set of SWRL rules that translate complex structures into lightweight relations ensures that both detailed and simplified knowledge representations coexist, facilitating better reasoning about, querying and interconnection of research data (RQ5).

In conclusion, the evolution of NFDIcore~1.0 to NFDIcore~2.0 demonstrates significant progress in addressing the challenges of representing and integrating heterogeneous research domains. By adopting foundational ontologies, integrating external standards and vocabularies, and enabling modular extensions, NFDIcore ensures a comprehensive representation, interoperability, and usability across various research disciplines. NFDIcore remains an ongoing development that will be refined and extended as more consortia are adapting the ontology. 
NFDIcore is a step towards interconnected research, set to offer new perspectives, interpretations, and visions on existing data, as well as to foster greater collaboration, novel discoveries and innovations in science and beyond. \\


\noindent\textbf{Acknowledgements.} This work is funded by Deutsche Forschungsgemeinschaft (DFG), project number 441958017.


\bibliographystyle{splncs04}

\end{document}